# University as catalyst of public R&D expenditures? An empirical assessment on EU NUTS 3 regions

Silvia Iossa
Institute of Management, Sant'Anna School of Advanced Studies, Pisa, Italy
silvia.iossa@santannapisa.it

Saverio Barabuffi
Institute of Management, Sant'Anna School of Advanced Studies,Pisa, Italy
L'EMbeDS Department, Sant'Anna School of Advanced Studies, Pisa, Italy
saverio.barabuffi@santannapisa.it

*Note*: Control function second stage estimates requires bootstrapped standard errors to be computed

## Abstract

This study examines whether universities shape the regional innovation effects of public R&D expenditure and whether this role varies across regional development trajectories. We analyse a panel of 1,166 EU NUTS 3 regions over the period 2007–2020, combining data on European public R&D funding, patenting performance, university presence, and regional development traps. To address the endogeneity of public R&D expenditure, we employ a control-function approach. The results show that public R&D expenditure positively affects patenting performance in both university and non-university regions, but the difference between them is not statistically significant once endogeneity is accounted for. University presence therefore does not systematically amplify the returns to public R&D. However, its moderating role becomes more pronounced in regions exposed to development-trap risk and in regions experiencing persistent stagnation, while evidence is weaker for trap intensity. These findings suggest that universities act as contingent rather than universal catalysts of public R&D. Their contribution appears particularly relevant in vulnerable and persistently stagnant regional trajectories, but less able to offset severe structural constraints. The study contributes to debates on the geography of public R&D and regional development traps by showing that the effectiveness of research funding depends not only on institutional presence, but also on the development trajectory in which that institutional capacity is embedded.

## 1. Introduction

Universities have long been recognised as important contributors to regional economic and innovation dynamics (Ferrara & Cavicchi, 2026). Beyond their traditional roles in education and research, they generate knowledge, develop human capital, support interactions among scientific, industrial, and public actors, and connect territories to wider knowledge networks (Bramwell & Wolfe, 2008; Breznitz & Feldman, 2012). Through these functions, universities can strengthen regional absorptive capacity and facilitate the recombination and application of knowledge (Buesa et al., 2010; Asheim et al., 2011). Their contribution may therefore extend beyond the direct production of knowledge to shaping the capacity of regions to translate research and innovation resources into tangible outcomes.

This role is particularly relevant in the context of public R&D expenditure. Public R&D represents a central policy instrument for fostering innovation, competitiveness, and economic development (Wanzenböck et al., 2020). At the European level, large-scale competitive funding schemes such as the EU Framework Programmes and Horizon 2020 have mobilised substantial resources to support collaborative research and innovation across countries and regions. Participation in these programmes provides access not only to financial resources, but also to transnational knowledge networks, research partnerships, and specialised capabilities. Yet their innovation outcomes remain highly uneven across territories (Fitjar, 2025). While linear models assumed a direct relationship between R&D inputs and innovation outputs, research increasingly shows that the effectiveness of R&D depends on the institutional, organisational, and knowledge conditions through which resources are absorbed and mobilised (Crescenzi & Giua, 2020). Universities may constitute one of these conditions, potentially influencing the extent to which public R&D expenditure is converted into regional innovation outcomes.

The importance of these institutional conditions becomes particularly apparent against the uneven geography of innovation and economic performance across the European Union (EU) (Iammarino et al., 2019). While some regions have successfully integrated into global knowledge networks and sustained high levels of growth, others have experienced prolonged stagnation in income, productivity, and employment. These divergent trajectories have also contributed to a "geography of discontent" in territories whose relative position and development prospects have deteriorated over time (Rodríguez-Pose, 2018). The concept of the regional development trap captures these dynamics by identifying regions that are unable to retain their economic dynamism while also underperforming their national and European peers (Diemer et al., 2022). Unlike classifications based solely on current income levels or structural disadvantage, regional development traps can emerge at different levels of development and through different trajectories. Their persistence reflects difficulties in adapting productive structures, developing new capabilities, and reconnecting with more dynamic development paths (MacKinnon et al., 2019).

Innovation is central to these dynamics. Development-trapped regions may receive innovation inputs without being able to translate them effectively into innovation outputs. Diemer et al. (2022), in particular, show that regions experiencing a development trap display weak growth in patenting performance, while differences in innovation inputs are less clearly associated with trapped status. In these contexts, the institutional and organisational capacities through which knowledge is absorbed, recombined, and applied may therefore become especially consequential. Universities may contribute to these capacities by supporting human capital formation, facilitating interactions among regional actors, and connecting territories to wider knowledge networks (Pinto, 2024). However, the presence of a university alone does not necessarily imply that public R&D resources will be translated more effectively into innovation outcomes, as the territorial effects of universities depend on their alignment with local capabilities and productive structures (Giraldi et al., 2026). Whether universities strengthen the relationship between public R&D expenditure and innovation outcomes, and whether their contribution becomes more important in regions experiencing a development trap, therefore remains an open empirical question.

Despite the established importance of universities in regional innovation processes (Tiffin & Kunc, 2011), existing research has paid limited attention to how they condition the regional innovation effects of competitive public R&D funding. Public R&D expenditure is frequently treated as a relatively homogeneous innovation input, although its outcomes depend on the institutional conditions available within recipient regions (Crescenzi & Giua, 2020). Similarly, while universities are widely recognised as important contributors to regional innovation, their role in shaping the effectiveness of public R&D expenditure remains insufficiently explored, particularly in regions experiencing a development trap. This paper addresses these gaps by analysing a panel of 1,166 EU NUTS 3 regions over the period 2007–2020. Specifically, it examines whether the presence of a university enhances the effect of public R&D expenditure on patenting performance and whether this effect is stronger in regions experiencing a regional development trap.

The paper makes two main contributions. First, it contributes to research on the territorial effects of public R&D by examining the institutional conditions under which research funding is translated into regional innovation outcomes (Crescenzi & Giua, 2020). Second, it advances the literature on universities and regional development by examining universities as potential catalysts of public R&D expenditure and assessing whether this role becomes particularly relevant in development-trapped regions (Bramwell & Wolfe, 2008; Breznitz & Feldman, 2012; Diemer et al., 2022).

The remainder of the paper is structured as follows. Section 2 reviews the relevant literature and develops the hypotheses. Section 3 presents the data and empirical strategy. Section 4 reports and discusses the results. Section 5 concludes by outlining the theoretical and policy implications.

## 2. Literature Background and Hypotheses

### 2.1 Universities and regional innovation capacity

Public R&D funding can support regional innovation by expanding the resources available for knowledge creation, collaboration, and technological development. Its effects, however, depend on the conditions under which these resources are mobilised. Evidence from European regions shows that participation in publicly funded R&D projects can contribute to the emergence of new technological specialisations, but that these effects vary according to the knowledge and technological capabilities already present in the region (Uhlbach et al., 2022). Public R&D expenditure should therefore not be regarded as an input producing homogeneous innovation effects across territories, but as a resource whose contribution depends on the capacity of regional actors to absorb and recombine the knowledge it generates.

In the European context, the EU Framework Programmes and subsequent Horizon programmes constitute major supranational instruments for supporting research and innovation. Funding is allocated competitively to collaborative projects involving universities, research organisations, firms, and other innovation actors located across European regions. These programmes therefore provide not only financial resources, but also opportunities to participate in transnational research and innovation networks and to access knowledge and capabilities distributed across organisational and geographical boundaries (Nepelski & Van Roy, 2021). However, participation alone does not determine the innovation outcomes generated by these resources. Their effects depend on how effectively funded knowledge and collaborations are absorbed and recombined within regional technological and institutional contexts.

Patenting provides a particularly relevant outcome through which to examine this relationship, as patents capture technological knowledge generated through inventive activity. Although Framework and Horizon programmes are not designed exclusively to generate patents, they support research and innovation activities that can produce technological knowledge, foster collaboration between scientific and industrial actors, and facilitate the recombination of knowledge across organisational and spatial boundaries. Research combining data on EU-funded projects, universities, firms, and patenting at the European NUTS 3 level reveals substantial heterogeneity in the relationship between academic and regional technological development (Caviggioli et al., 2023). This suggests that the availability of research resources alone may be insufficient to generate technological outputs, and that their translation into patenting activity also depends on the capacity of regional actors to exploit and connect funded knowledge to existing technological capabilities.

Universities can play an important role in this process. They provide scientific knowledge and research capabilities and can create connections between academic research and industrial innovation. Evidence on the co-evolution of university and regional technological specialisation shows that universities may contribute to the development of local technological capabilities, although the strength and direction of this relationship vary across regions and according to university characteristics (Caviggioli et al., 2023). Research on university–industry relationships similarly indicates that the exploitation and commercialisation of academic knowledge depend on the ability to bridge technological and cognitive distances between universities and firms (Aksoy et al., 2022). These mechanisms can strengthen the regional capacity to absorb, connect, and exploit knowledge generated through R&D investment.

Universities may therefore influence not only the direct production of knowledge but also the extent to which externally funded research resources are converted into technological outputs. In this sense, their presence may strengthen the institutional and knowledge conditions through which public R&D expenditure contributes to regional patenting performance. Accordingly:

*H1. The presence of a university in EU NUTS 3 regions enhances the positive effect of public R&D expenditure on patenting performance.*

### 2.2 Regional development traps and the role of universities

The capacity of public R&D expenditure to generate innovation outcomes is likely to vary across regional development trajectories. Regions differ substantially in their ability to mobilise knowledge, human capital, networks, and institutional resources for innovation, and these differences may become particularly consequential where economic and technological development has stagnated. Evidence from European regions shows that the combinations of conditions supporting innovation differ across levels of regional development, with public R&D, human capital, and collaborative networks playing different roles in economically developed and lagging regions (Filippopoulos & Fotopoulos, 2022). The concept of the regional development trap provides a dynamic perspective on these differences. Diemer et al. (2022) define a region as development trapped when it is unable to retain its economic dynamism in terms of income, productivity, and employment while simultaneously underperforming its national and European peers. Regional development traps are therefore not confined to low-income or peripheral regions but may emerge at different levels of development and through different trajectories. Their defining feature is a persistent deterioration in economic dynamism relative both to a region’s own past performance and to relevant national and European benchmarks. Recent research further shows that development traps differ in their risk, intensity, and duration (Rodríguez-Pose et al., 2024). Risk captures the extent to which a region displays the conditions associated with falling into a development trap; intensity reflects the severity of the deterioration in its development trajectory; and duration captures the persistence of the trapped condition over time. These dimensions highlight that regional development traps are heterogeneous: regions may differ not only in whether they experience stagnation, but also in how severely and persistently their development trajectory deteriorates.

Innovation represents an important component of these dynamics. Development-trapped regions display particularly weak growth in patenting performance, whereas differences in the growth of innovation inputs, including R&D expenditure and science and technology personnel, are less clearly associated with trapped status (Diemer et al., 2022). This suggests that the challenge faced by these regions concerns not only access to innovation resources, but also their capacity to translate those resources into technological outputs. Where existing capabilities and connections are weaker, additional R&D expenditure may therefore require stronger institutional and organisational capacities to generate innovation outcomes. Universities may be particularly relevant in this process. Their research capabilities, human capital, and connections with firms and other innovation actors can contribute to the absorption and recombination of knowledge and to the mobilisation of resources around regional innovation activities. Research on universities operating in structurally challenging regional contexts also suggests that their capacity to contribute to regional change depends on institutional capacities and on their ability to engage with the surrounding regional environment (Medina-Bueno et al., 2024). Universities may therefore provide some of the knowledge and institutional capacities that become especially valuable when established regional development trajectories have weakened. This does not imply that university presence is sufficient to enable a region to escape a development trap. Development traps encompass income, productivity, and employment dynamics that extend beyond technological innovation. Rather, the argument concerns a more specific mechanism: the extent to which universities condition the conversion of public R&D expenditure into technological outputs. If development-trapped regions face greater difficulties in translating innovation inputs into outputs, the institutional and knowledge capacities associated with universities may become particularly consequential in these contexts. Accordingly:
*H2. The positive effect of university presence on the relationship between public R&D expenditure and patenting performance is stronger in EU NUTS 3 regions experiencing a regional development trap.*

## 3. Methodology and data

### *3.1. Methodology*

First, to empirically investigate the effect of the presence of a university in NUTS 3 regions' returns to scale to public R&D expenditures (H1), we estimate the following empirical model:

$$Y_{i,t} = \alpha + \beta_1 Public\ R\&D\ expenditures_{i,t-1} \times University_{i,t-1} + X'_{i,t-1}\beta + \eta_i + \iota_t + \varepsilon_{i,t}$$

*Eq. 1*

where $Y$ represents the output of regional patenting performance, $Public\ R\&D\ expenditures \times University$ measures the interaction term between $Public\ R\&D\ expenditures$ and the dummy variable $University$ which indicates whether the NUTS 3 region $i$ hosts a university. Then, $X'$ is a vector of control variables, while $\eta$ and $\iota$ represent regional and time fixed effects to sort out regional time-invariant characteristics and potential trends existing in a specific year, respectively. Lastly $\varepsilon$ accounts for the idiosyncratic errors.

Second, to study the effect of the presence of university of NUTS 3 in left-behind NUTS 3 regions' returns to scale to public R&D expenditure (H2), we augment Eq. 1 by interacting the variable $Left\ behind$ with $Public\ R\&D\ expenditures \times University$, estimating the following:

$$Y_{i,t} = \alpha + \beta_1 Public\ R\&D\ expenditures_{i,t-1} \times University_i \times Development\ Trap_i + X'_{i,t-1}\beta + \eta_i + \iota_t + \varepsilon_{i,t}$$

*Eq. 2*

where $Development\ Trap$ is a dummy variable that indicates whether the NUTS 3 region $i$ is experiencing a regional development trap in the period of analysis, or not.

However, *Public R&D expenditure* may be endogenous to regional patenting performance. Reverse causality could arise if regions with stronger innovative performance subsequently attract greater public R&D funding, while unobserved regional characteristics correlated with both innovation capacity and the allocation of public R&D may further bias the OLS estimates of *Eq. 1* and *Eq. 2*. To address this concern, we complement our baseline estimates with a control function (CF) approach (Wooldridge, 2015), which is particularly suited to our setting since the potentially endogenous variable enters the model interacted with university presence (Eq. 1) and, with both university presence and development trap status (Eq. 2).

The CF approach proceeds in two steps. In the first stage, we regress *Public R&D expenditures* on an instrument, together with the same set of control variables and fixed effects included in the main specification:

$$Public\ R\&D\ expenditures_{i,t-1} = \alpha + \beta_1 IV\ Public\ R\&D\ expenditures_{i,t-1} \times University_i + X'_{i,t-1}\beta + \eta_i + \iota_t + \upsilon_{i,t-1}$$

*Eq. 3*

and

$$Public\ R\&D\ expenditures_{i,t-1} = \alpha + \beta_1 IV\ Public\ R\&D\ expenditures_{i,t-1} \times University_i \times Development\ Trap_i + X'_{i,t-1}\beta + \eta_i + \iota_t + \upsilon_{i,t-1}$$

*Eq. 4*

where the instrumental variable is interacted with $University_i$ for Eq.1, and additionally with $Regional\ Development\ Trap_i$ for the specification in Eq. 2, to allow the instrument to vary across subgroups defined by these two exogenous indicators. The first-stage specification is therefore designed to mirror this structure, ensuring that each subgroup-specific coefficient of interest is identified by its own source of exogenous variation. From the first-stage regressions we recover the residual $\upsilon_{i,t-1}$, which captures the component of *Public R&D expenditures* left unexplained by the instrument and controls, and which is potentially correlated with the error term of *Eq. 1* and *Eq. 2*.

The second stages estimate Eq. 1 and Eq. 2, respectively, including both the endogenous variable interacted with the same subgroup indicators used in the corresponding main specification, and the first-stage residual $\upsilon_{i,t-1}$:

$$Y_{i,t} = \alpha + \beta_1 Public\ R\&D\ expenditures_{i,t-1} \times University_{i,t-1} + X'_{i,t-1}\beta + \eta_i + \iota_t + \upsilon_{i,t-1} + \varepsilon_{i,t}$$

*Eq. 5*

and

$$Y_{i,t} = \alpha + \beta_1 Public\ R\&D\ expenditures_{i,t-1} \times University_i \times Development\ Trap_i + X'_{i,t-1}\beta + \eta_i + \iota_t + \upsilon_{i,t-1} + \varepsilon_{i,t}$$

*Eq. 6*

Including the first-stage residual purges the component of public R&D expenditure that is correlated with the error term from the estimated effect, following the logic of the two-stage residual inclusion method (Wooldridge, 2015).

### *3.2. Data*

#### Dependent variable

We proxied NUTS 3 regions' patenting performance in terms of patents applications to the European Patent Office (EPO) during the period 2007-2020. We use inventors' addresses to more accurately reflect the knowledge-creation process, and we employ a fractional counting method to proportionally credit the inventive contribution to each NUTS 3 region. Then, we express our dependent variable in terms of number of patents per capita to control for the different sizes of each NUTS 3 region.

#### Independent variable

We measured the $Public\ R\&D\ expenditures$ by leveraging the RISIS EUPRO dataset (Heller-Schuh et al., 2025) that collects information on project based R&D collaboration networks in Europe constituted under the European Framework Programme 7 (FP7) and Horizon 2020 (H2020). We collected the information related to projects' contribution of the EU associated to each participant, and their duration. Then, the EU contribution has been allocated pro-rata over each projects' duration and aggregated at NUTS 3 level. Finally, we assume that $Public\ R\&D\ expenditures$ cumulate with an obsolescence rate represented by the discount procedure according to the Perpetual Inventory Method (PIM) suggested by OECD (2009), taking an average discount rate *d* of 15%, and rescaled for GDP to control for the different economic size of each NUTS 3 region. The final formulation for our main regressor of interest for NUTS 3 region *i* at time *t* is the following:

$$Public\ R\&D\ expenditures_{i,t} = \frac{\sum_{s=0}^{t}\left(FP7\ \&\ H2020\ EU\ contribution_{i,s} \times e^{-d(t-s)}\right)}{GDP_{i,t}}$$

*Eq. 3*

#### Instrumental variable

To address the potential endogeneity of public R&D expenditure with respect to regional patenting performance, we construct a shift-share instrumental variable (Goldsmith-Pinkman et al., 2020) that exploits each region's pre-determined position in the FP6 collaboration network, interacted with the subsequent stock of FP7 and H2020 funding to its network partners. Specifically, for each NUTS3 region $i$, we compute Bonacich centrality (Bonacich, 1987) within the 2002-2006 FP6 regional collaboration network, derived from bilateral co-participation ties recorded in FP6 consortium projects prior to the start of our sample period. This centrality measure captures not only the number of direct research linkages a region held during FP6, but also the extent to which its partners were themselves well-connected, thereby weighting each tie by the broader importance of the partner within the network (Ancona *et al*., 2024). We then interact this pre-period centrality with the stock of FP7 and H2020 expenditure received by region *i*'s FP6 network partners, excluding the top decile of its own FP6 partners when constructing the partner-funding shock. This partner-specific trimming mitigates concerns that the instrument's identifying variation is driven by each region's most frequent collaborators.

### Interaction terms

The variable $University$ is a time-invariant dummy variable that takes the value 1 if the NUTS 3 region $i$ hosts a university headquarter, and 0 otherwise. We focus on university headquarters rather than satellite campuses because headquarters concentrate core governance, research infrastructures, and the majority of academic staff, representing the main institutional locus of scientific production and research collaboration within regional innovation systems. Satellite campuses, in contrast, often host more limited teaching or administrative activities and do not necessarily reflect the full research capacity of the institution.

To operationalize $Development\ Trap$ we followed Diemer *et al.* (2022) and Rodriguez-Pose *et al.* (2024). Firstly, we computed $a_{i,t}^{r}$ as the deviation from the regional growth rate of region $i$, $a_{i,t}^{C}$as the deviation from the national rate of the country $c$ where NUTS 3 region $i$ is located, and $a_{i,t}^{EU}$ as the deviation from the European union growth rate as follows:

$$a_{i,t}^{R} = g_{i,t,t-n} - g_{i,t,t-2n}$$

*Eq. 3*

$$a_{i,t}^{C} = g_{i,t,t-n} - g_{c,t,t-n}$$

*Eq. 4*

$$a_{i,t}^{EU} = g_{i,t,t-n} - g_{EU,t,t-n}$$

*Eq. 6*

where $g_{i,t,t-n}$, $g_{c,t,t-n}$, and $g_{EU,t,t-n}$ are the average compound annual growth rate for NUTS 3 region $i$, the average compound annual growth rate for country $c$ in which the NUTS 3 region $i$ is located, and the average compound annual growth rate for the EU, respectively. While $n$ is set equal to 5.
We computed the average compound annual growth rate as described below:

$$g_{k,t,t-n} = n^{-1} \times ln \frac{z_{k,t}}{z_{k,t-n}}$$

*Eq. 5*

where *k* represents NUTS 3 region *i*, country *c*, or *EU*, while z takes the value of *GDP per capita*, *labour productivity* or *employment to population ratio*, and n is set equal to 5. Then, we define a set of 9 dummy variables[1] that takes the value 1 for positive accelerations, and 0 otherwise.
As proposed by Rodriguez-Pose *et al.* (2024), we developed three indices that capture Regional Development Trap: *risk*[2], *intensity*[3], and *duration*.
*Risk* captures the Regional Development Trap risk, and is computed by adding the 9 dummy variables, and rescaling the sum to a score between 0 and 1, Consequently, we define the NUTS 3 regions experiencing a Regional Development Trap as those with *Risk* mean scores above the 75th percentile throughout the series. Then, *intensity* captures Regional Development Trap intensity, and is computed by averaging the acceleration, and standardize it by using the mean and standard deviations of the first year in the series' distribution (i.e.: 2007). As previously, we define the NUTS 3 regions experiencing a Regional Development Trap as those with *DT2* mean scores above the 75th percentile throughout the series. Finally, we compute the *duration* by counting the years a region has been trapped, hence having a DT1 above 0.5 in a given year. We then define NUTS 3 regions as experiencing a Regional Development trap if *duration is* above the 75th percentile throughout the series.

### Control variables

As a set of control variables we use NUTS 3 level variables such as Gross Domestic Product (GDP) per capita, employment-population ratio, investment-labour ratio measured as gross fixed capital formation divided by hours worked, the sectoral share of regional GVA with respect to agriculture, industry, construction, market services, and nonmarket services. The economic variables are then expressed in EUR2015 at constant prices and transformed by using the sine hyperbolic function.

Table 1 reports the data sources, frequency distribution and descriptive statistics, while Table 2 shows the correlation matrix.

[INSERT TABLE 1 HERE]

[INSERT TABLE 2 HERE]

## 4. Results

Table 3 reports the OLS and CF estimation results of the effect of between public R&D expenditure, presence of universities and patenting performance. Column 1 shows a positive and significant association between public R&D expenditure and patenting ($\beta = 0.0856$, $p < 0.01$), effect that holds once regional controls are included in column 2 ($\beta = 0.0744$, $p < 0.01$). Column 3 splits this effect by university presence, showing that the coefficient is small and non-statistically significant in regions without a university ($\beta = 0.0310$, $p > 0.10$), while it is positive and significant in university-hosting regions ($\beta = 0.0765$, $p < 0.01$). This pattern is consistent with an absorptive-capacity account in which local research infrastructure is necessary to convert public R&D funding into patentable innovation, and provides initial support for H1. This pattern is consistent with regional absorptive-capacity, in which only regions with sufficient local research infrastructure are able to translate additional public R&D funding into patentable innovation.
Then, column 4 reports the CF estimates, obtained by including the first-stage residual from our centrality-based shift-share instrument as an additional regressor. The first-stage F-statistics for the excluded instrument indicate that the instrument is not weak. Once endogeneity is addressed, both coefficients increase substantially and become statistically significant ($\beta$ =0.3249, $p < 0.05$ for non-university regions; $\beta$ = 0.3580, $p < 0.05$ for university regions). The coefficient on the residual term is statistically significant ($\beta$ =-0.2858, $p < 0.05$), confirming that public R&D expenditure was indeed endogenous and the correction suggest attenuation bias from measurement error in public R&D expenditure. FP7 and H2020 funding are formally awarded through competitive, merit-based calls, the applied excellence criteria evaluate scientific and proposal quality rather than a region's capacity to translate research into patentable output, which can generate a negative region-level correlation between existing patenting capacity and funding received, biasing the OLS coefficient downward relative to the endogeneity-corrected estimate. A formal test of equality between the two control-function coefficients fails to reject the null hypothesis that the effect of public R&D expenditure on patenting is the same in university and non-university regions (difference = -0.060, $p >0.10$). This confirms that, once the endogeneity of public R&D expenditure is addressed, the two subgroup effects are statistically indistinguishable from one another.

[1] Three dummy for each acceleration for the three variables defined by Diemer *et al.* (2022).
[2] Diemer et al. (2022) define the variable *risk* as *DT1*
[3] Diemer et al. (2022) refer to *intensity* as *DT2*.

This result qualifies expectations derived from previous research. Existing studies suggest that the innovation effects of public R&D depend on complementary regional conditions rather than on funding alone. Filippopoulos and Fotopoulos (2022), for example, show that public R&D contributes to regional innovation through configurations that combine R&D resources with other regional capabilities, including human capital and collaboration networks. Their findings further indicate that, in developed European regions, public R&D frequently forms part of broader configurations involving internal capabilities, networks of collaboration, and access to external knowledge. Evidence on EU Framework Programmes similarly suggests that the innovation effects of public research support can be stronger when firms collaborate with universities (Szücs, 2018). Against this background, the absence of a statistically significant difference between university and non-university regions suggests that the mere presence of a university may be insufficient to capture the complementarities through which public R&D is converted into technological outputs. Universities can provide scientific knowledge, human capital, research infrastructure, and connections to wider knowledge networks, but these functions may matter only when they are effectively mobilised and connected to other regional actors and capabilities. The results therefore point towards a more differentiated interpretation of universities as institutional catalysts: their contribution to the returns of public R&D may depend less on territorial presence per se than on the specific relationships and capabilities through which university knowledge becomes embedded in regional innovation processes. Rather than questioning the potential contribution of universities to regional innovation, the findings suggest that university presence alone is too coarse a condition to explain differences in the effectiveness of public R&D expenditure across regions.

[INSERT TABLE 3 HERE]

Table 4 reports the estimation results of the interaction between Public R&D Expenditures, presence of university and regional Development Trap. The triple-interaction OLS estimates reveal heterogeneous patterns across the three development-trap dimensions. In column 1 the effect of risk is positive and statistically significant in the university/at-risk regions ($\beta = 0.1637$, $p < 0.05$) and in the university/not-at-risk regions ($\beta = 0.0401$, $p < 0.10$), with both non-university cells non statistically significant. Column 3 shows the effect for intensity which is positive and statistically significant in the non-university/intensity regions ($\beta = 0.1546$, $p < 0.10$) and university/no intensity regions ($\beta = 0.0913$, $p < 0.01$), but negative and non-statistically significant in the university/intensity regions ($\beta = -0.0124$, $p>0.1$). Column 5 reports the effect of duration, showing positive and statistically significant effects for duration-trapped regions (non-university: $\beta = 0.1007$, $p < 0.05$; university: $\beta = 0.1715$, $p < 0.05$), with both short-duration regions statistically indistinguishable from zero.

Then, columns 2, 4, and 6 report the CF-corrected estimates for each dimension. The first-stage F-statistics confirm the instrument remains relevant across all subgroups. Once endogeneity is addressed, we find that regions with the presence of university have a stronger effect on patenting performance when they experience regional development trap *risk* and *duration* ($\beta = 0.4064$, $p < 0.01$, and $\beta = 0.4131$, $p < 0.01$, respectively), while it is not reflected with *intensity* ($\beta = 0.2665$, $p < 0.01$). The results provide differentiated support for H2. The moderating role of university presence is consistent with the hypothesis when regional development traps are captured in terms of risk and duration, whereas the evidence is weaker and less consistent when considering trap intensity. This suggests that the role of universities varies not simply according to whether a region is development trapped, but according to the nature of the development trajectory in which it is embedded.

This pattern can first be interpreted from a regional path development perspective. Regional economies evolve through historically accumulated productive, knowledge, and institutional structures that shape subsequent development opportunities and may generate self-reinforcing trajectories (MacKinnon et al., 2019). Regional development traps emerge when these trajectories lose dynamism and regions experience persistent difficulties in adapting their productive structures and developing new capabilities (Diemer et al., 2022). In such contexts, regional renewal requires not only additional resources, but also the capacity to mobilise and recombine knowledge in ways that broaden existing development possibilities.

Universities may be particularly relevant to these processes because they provide knowledge and capabilities that are not entirely dependent on the existing productive trajectory of a region. Through research activities, human capital formation, and connections to wider knowledge networks, universities can facilitate access to extra-local knowledge and support the recombination of existing and new capabilities. These functions are consistent with accounts of path creation that emphasise agency and knowledge recombination in opening alternative development trajectories (Garud et al., 2010; MacKinnon et al., 2019). The stronger moderating effect observed for development-trap risk and duration is consistent with this interpretation. In regions exposed to increasing vulnerability, university-related capabilities may broaden the knowledge resources available before weaker trajectories become more deeply entrenched. Similarly, where stagnation persists over time, the relatively stable knowledge infrastructure associated with universities may provide resources for experimentation and recombination that are difficult to generate exclusively from the existing regional productive base.

The weaker evidence for development-trap intensity, however, points to an important boundary condition. Balland and Boschma (2025) argue that regional traps should not be understood as uniform states: regions can be trapped to a greater or lesser extent, depending on the capabilities available to adapt, diversify, and develop new activities. Their evolutionary perspective highlights that regional development opportunities are conditioned by capabilities accumulated in people, institutions, and economic activities, which determine the trajectories that remain accessible to a region. Where these capabilities are particularly weak, moving towards new activities may require coordinated investment in education, research, firms, and institutions, together with stronger connections to external sources of complementary knowledge. The weaker moderating effect observed for trap intensity may therefore suggest that university presence alone has more limited leverage when regional structural constraints are particularly severe.

This interpretation does not imply that universities enable regions to escape development traps. Rather, they may contribute to one specific mechanism relevant to regional renewal: strengthening the capacity to mobilise public R&D towards new technological outputs. Their contribution appears particularly relevant where regions face the risk of becoming trapped or have experienced persistent stagnation, while becoming less evident as the severity of the trap increases. H2 is therefore supported in a differentiated rather than uniform way, suggesting that the catalytic role of universities depends on the nature and depth of the regional development trajectory in which they operate.

[INSERT TABLE 4 HERE]

## 5. Conclusions

This study examines whether the presence of universities shapes the relationship between public R&D expenditure and patenting performance across European NUTS 3 regions, and whether this role becomes more important in regions experiencing a regional development trap.
The results provide a more differentiated picture than initially expected. Once the endogeneity of public R&D expenditure is addressed, university presence does not generate a statistically distinct moderating effect on the relationship between public R&D and patenting performance. H1 is therefore not supported. This suggests that the mere presence of a university is not sufficient to explain why similar levels of public R&D expenditure generate different technological outcomes across regions. At the same time, the results for regional development traps reveal greater heterogeneity. University presence strengthens the relationship between public R&D expenditure and patenting performance when development traps are captured in terms of risk and duration, while the evidence is weaker for trap intensity. H2 therefore receives differentiated support.

These findings contribute to the literature in two main ways. First, they qualify the view of universities as general institutional amplifiers of public R&D. Previous research shows that the innovation effects of public R&D depend on complementary regional conditions, including human capital, collaboration networks, and other capabilities through which knowledge can be absorbed and recombined (Filippopoulos & Fotopoulos, 2022). Universities may contribute to these conditions through knowledge production, human capital formation, and connections with wider innovation networks, but the results suggest that these functions do not translate automatically into higher returns to public R&D simply because a university is located in the region. Their contribution is therefore likely to depend on the specific capabilities and relationships through which academic knowledge becomes connected to the surrounding regional economy.

Second, the findings show that the role of universities becomes more consequential when regional development trajectories are taken into account. Regional development traps are characterised by persistent difficulties in sustaining economic dynamism and developing new capabilities (Diemer et al., 2022), while evolutionary perspectives emphasise that the opportunities available to regions depend on historically accumulated capabilities embodied in people, institutions, and economic activities (MacKinnon et al., 2019; Balland & Boschma, 2025). The stronger effects observed for development-trap risk and duration suggest that universities may contribute to the mobilisation of knowledge and research resources in regions facing vulnerable or persistently stagnant trajectories. However, the weaker evidence for trap intensity indicates that university presence alone may have limited leverage where structural constraints are particularly severe.
This latter result is particularly relevant considering recent evolutionary approaches to regional development traps. Balland and Boschma (2025) stress that regions may be trapped to different degrees and that severe structural constraints can substantially restrict their ability to diversify and develop new activities. Where local capabilities are particularly weak, regional renewal may require coordinated investments in research, education, firms, institutions, and external connections rather than reliance on a single institutional actor. Our findings are consistent with this interpretation: universities may act as catalysts under some development conditions, but they cannot be expected to compensate on their own for deeply rooted structural weaknesses.

The findings also have implications for innovation policy. Public R&D funding should not be expected to generate homogeneous innovation outcomes across territories, as its effects depend on the capabilities and institutional conditions through which resources are mobilised (Fitjar, 2025; Filippopoulos & Fotopoulos, 2022). This reinforces the relevance of place-sensitive approaches to R&D policy. At the same time, the results caution against assuming that the presence of a university is sufficient to compensate for weak regional innovation capacities. Policies aimed at strengthening the territorial contribution of universities should therefore focus not only on institutional presence, but also on the relationships, capabilities, and knowledge connections through which universities interact with regional actors (Pinto, 2024).

The differentiated results across development-trap dimensions further suggest that policy responses should consider the nature of the development trajectory faced by individual regions. Regions at risk of becoming trapped, regions experiencing persistent stagnation, and regions characterised by more severe deterioration may require different combinations of research support and capability-building interventions. This is consistent with the broader argument that regional diversification opportunities depend on existing capabilities and that regions facing deeper structural constraints may require more coordinated and institutionally demanding interventions (Balland & Boschma, 2025).

Several limitations should be acknowledged. First, the analysis focuses on patenting and therefore captures technological innovation rather than the broader range of innovation outcomes that universities and public R&D may generate. This limitation is particularly relevant for weaker regions, where innovation may rely less strongly on patenting and more on other forms of innovative activity (Filippopoulos & Fotopoulos, 2022). Second, the measure of public R&D expenditure focuses on funding allocated through European Framework Programmes and does not capture other forms of public or private R&D investment.

Future research could therefore investigate complementarities between European research funding, private R&D expenditure, and Cohesion Policy resources, as well as the role of interregional research collaborations in connecting development-trapped regions to external capabilities. This latter direction appears particularly relevant given evidence that access to complementary capabilities outside the region can support adaptation and diversification (Balland & Boschma, 2021; Balland & Boschma, 2025). Longitudinal research could investigate whether and through which mechanisms university-related capabilities contribute to changes in regional development trajectories over time.

## References

Aguiar, L., & Gagnepain, P. (2017). European cooperative R&D and firm performance: Evidence based on funding differences in key actions. *International Journal of Industrial Organization, 53*, 1–31.

Aksoy, A. Y., Pulizzotto, D., & Beaudry, C. (2022). University-industry partnerships in the smart specialisation era. *Technological Forecasting and Social Change, 176*, 121438. doi:10.1016/j.techfore.2021.121438.

Ancona, A., Cerqueti, R., & Grassi, R. (2024). How do partner selection strategies affect the amount of funding in collaborative research projects? Evidence using the dual-projection approach. *Technological Forecasting and Social Change, 208*, 123744. doi:10.1016/j.techfore.2024.123744.

Asheim, B. T., Smith, H. L., & Oughton, C. (2011). Regional innovation systems: Theory, empirics and policy. *Regional Studies, 45*(7), 875–891. doi:10.1080/00343404.2011.596701.

Audretsch, D. B., & Feldman, M. P. (1996). R&D spillovers and the geography of innovation and production. *American Economic Review, 86*(3), 630–640.

Balland, P.-A., & Boschma, R. (2021). Complementary interregional linkages and Smart Specialisation: An empirical study on European regions. *Regional Studies, 55*(6), 1059–1070. doi:10.1080/00343404.2020.1861240.

Balland, P.-A., & Boschma, R. (2025). An evolutionary approach to regional development traps in Europe. *Regional Studies, 59*(1), 2586708. doi:10.1080/00343404.2025.2586708.

Balland, P.-A., Jara-Figueroa, C., Petralia, S., Steijn, M., Rigby, D., & Hidalgo, C. A. (2020). Complex economic activities concentrate in large cities. *Nature Human Behaviour, 4*, 248–254.

Becker, B. (2015). Public R&D policies and private R&D investment: A survey of the empirical evidence. *Journal of Economic Surveys, 29*(5), 917–942. doi:10.1111/joes.12074.

Benneworth, P. (Ed.). (2018). *Universities and regional economic development*. Routledge.

Blankenberg, A. K., & Buenstorf, G. (2016). Regional co-evolution of firm population, innovation and public research? Evidence from the West German laser industry. *Research Policy, 45*(4), 857–868.

Bol, T., de Vaan, M., & van de Rijt, A. (2018). The Matthew effect in science funding. *Proceedings of the National Academy of Sciences, 115*(19), 4887–4890.

Bonacich, P. (1987). Power and centrality: A family of measures. *American Journal of Sociology, 92*(5), 1170–1182. doi:10.1086/228631.

Bonitz, M., Bruckner, E., & Scharnhorst, A. (1997). Characteristics and impact of the Matthew effect for countries. *Scientometrics, 40*, 407–422.

Boschma, R. A., & Lambooy, J. G. (1999). Evolutionary economics and economic geography. *Journal of Evolutionary Economics, 9*(4), 411–429. doi:10.1007/s001910050089.

Boschma, R., Fitjar, R. D., Giuliani, E., & Iammarino, S. (2025). Unseen costs: The inequities of the geography of innovation. *Regional Studies*. doi:10.1080/00343404.2024.2445594.

Bramwell, A., & Wolfe, D. A. (2008). Universities and regional economic development: The entrepreneurial University of Waterloo. *Research Policy, 37*(8), 1175–1187.

Breschi, S., & Cusmano, L. (2006). Unveiling the texture of a European Research Area: Emergence of oligarchic networks under EU Framework Programmes. In *Knowledge flows in European industry* (pp. 294–324). Routledge.

Breznitz, S. M., & Feldman, M. P. (2012). The engaged university. *The Journal of Technology Transfer, 37*(2), 139–157.

Broekel, T. (2015). Do cooperative research and development (R&D) subsidies stimulate regional innovation efficiency? Evidence from Germany. *Regional Studies, 49*(7), 1087–1110.

Buesa, M., Heijs, J., & Baumert, T. (2010). The determinants of regional innovation in Europe: A combined factorial and regression knowledge production function approach. *Research Policy, 39*(6), 722–735.

Calignano, G., & Nordli, A. J. (2024). Less-advanced regions in EU innovation networks: Could nanotechnology represent a possible trigger for path upgrading? *PLOS ONE, 19*(1), e0288669. doi:10.1371/journal.pone.0288669.

Caviggioli, F., Colombelli, A., De Marco, A., Scellato, G., & Ughetto, E. (2023). Co-evolution patterns of university patenting and technological specialization in European regions. *The Journal of Technology Transfer, 48*, 216–239. doi:10.1007/s10961-021-09910-0.

Cinar, R., & Benneworth, P. (2020). Why do universities have little systemic impact with social innovation? An institutional logics perspective. *Growth and Change, 52*(2), 751–769. doi:10.1111/grow.12367.

Cooke, P., Gómez Uranga, M., & Etxebarria, G. (1997). Regional innovation systems: Institutional and organizational dimensions. *Research Policy, 26*, 475–491.

Crescenzi, R., & Giua, M. (2020). One or many cohesion policies of the European Union? *Regional Studies, 54*(1), 10–20.

David, P. A., Hall, B. H., & Toole, A. A. (2000). Is public R&D a complement or substitute for private R&D? *Research Policy, 29*(4–5), 497–529.

Diemer, A., Iammarino, S., Rodríguez-Pose, A., & Storper, M. (2022). The regional development trap in Europe. *Economic Geography, 98*(5), 487–509.

Doloreux, D., & Porto Gomez, I. (2017). A review of (almost) 20 years of regional innovation systems research. *European Planning Studies, 25*(3), 371–387.

Duranton, G., & Puga, D. (2004). Micro-foundations of urban agglomeration economies. In J. V. Henderson & J. F. Thisse (Eds.), *Handbook of regional and urban economics* (pp. 2063–2117).

Eder, J. (2019). Innovation in the periphery: A critical survey and research agenda. *International Regional Science Review, 42*, 119–146.

Feldman, M., & Kogler, D. (2010). Stylized facts in the geography of innovation. In B. H. Hall & N. Rosenberg (Eds.), *Handbook of the economics of innovation* (Vol. 1, pp. 381–410).

Fernández-Esquinas, M., & Pinto, H. (2014). The role of universities in urban regeneration. *European Planning Studies, 22*(7), 1462–1483.

Ferrara, C., & Cavicchi, A. (2026). The quadruple helix and the role of HEIs as innovation brokers in disadvantaged areas. *Regional Studies, 60*(1), 2594501. doi:10.1080/00343404.2025.2594501.

Filippopoulos, N., & Fotopoulos, G. (2022). Innovation in economically developed and lagging European regions: A configurational analysis. *Research Policy, 51*(2), 104424. doi:10.1016/j.respol.2021.104424.

Fitjar, R. D. (2025). Does public R&D funding reinforce regional disparities? *Research Policy, 54*(9), 105312.

Forth, T., & Jones, R. (2020). *The missing £4 billion: Making R&D work for the whole UK*. Nesta.

Garud, R., Kumaraswamy, A., & Karnøe, P. (2010). Path dependence or path creation? *Journal of Management Studies, 47*(4), 760–774.

Giraldi, L., Rossi, L., & Landoni, M. (2026). Technological proximity and place-based innovation policy outcomes: University–industry alignment in Italian provinces (2000–2022). *European Planning Studies*, 1–23. doi:10.1080/09654313.2026.2727517.

Goddard, J., Hazelkorn, E., Kempton, L., & Vallance, P. (2016). *The civic university*. Edward Elgar.

Goldsmith-Pinkham, P., Sorkin, I., & Swift, H. (2020). Bartik instruments: What, when, why, and how. *American Economic Review, 110*(8), 2586–2624. doi:10.1257/aer.20181047.

Heller-Schuh, B., Barber, M., Bilalli Shkodra, X., Scherngell, T., Wolfmayr, A. K., & Zahradnik, G. (2025). *Documentation of RISIS datasets: EUPRO*. Zenodo. doi:10.5281/zenodo.18233449.

Iacobucci, D., & Perugini, F. (2023). Innovation performance in traditional industries. *Technological Forecasting and Social Change, 189*, 122340.

Iammarino, S., Rodríguez-Pose, A., & Storper, M. (2019). Regional inequality in Europe. *Journal of Economic Geography, 19*(2), 273–298.

Kemeny, T., & Storper, M. (2020). *Superstar cities and left-behind places: Disruptive innovation, labor demand, and interregional inequality*. International Inequalities Institute Working Paper No. 41, London School of Economics and Political Science.

MacKinnon, D., Dawley, S., Pike, A., & Cumbers, A. (2019). Rethinking path creation. *Economic Geography, 95*(2), 113–135.

Martin, R., Martinelli, F., & Clifton, J. (2022). Rethinking spatial policy. *Cambridge Journal of Regions, Economy and Society, 15*(1), 3–21.

Mayer, H., & Motoyama, Y. (2020). Entrepreneurship in small towns. *Entrepreneurship & Regional Development, 32*(7–8), 467–472.

McCann, P., & Ortega-Argilés, R. (2013). Modern regional innovation policy. *Cambridge Journal of Regions, Economy and Society, 6*(1), 1–30.

Moreno, R., Paci, R., & Usai, S. (2005). Spatial spillovers and innovation activity in European regions. *Environment and Planning A, 37*(10), 1793–1812. doi:10.1068/a37341.

Morgan, K. (1997). The learning region. *Regional Studies, 31*, 491–503.

Nepelski, D., & Van Roy, V. (2021). Innovation and innovator assessment in R&I ecosystems: The case of the EU Framework Programme. *The Journal of Technology Transfer, 46*, 792–827. doi:10.1007/s10961-020-09814-5.

Neuländtner, M., & Scherngell, T. (2020). What drives R&D collaboration networks? *The Annals of Regional Science, 65*, 743–773.

OECD. (2009). The perpetual inventory method – Overview. In *Measuring capital: OECD manual* (2nd ed., pp. 87–90). OECD Publishing.

Pinto, H. (2024). Universities and institutionalization of regional innovation policy in peripheral regions: Insights from the smart specialization in Portugal. *Regional Science Policy & Practice, 16*(1), 12659. doi:10.1111/rsp3.12659.

Pinto, H., Uyarra, E., & Fernández-Esquinas, M. (2018). University roles in a peripheral region. In P. Benneworth (Ed.), *Universities and regional economic development*. Routledge.

Pugh, R. (2017). Universities and economic development in lagging regions: "Triple helix" policy in Wales. *Regional Studies, 51*(7), 982–993.

Radosevic, S., & Stancova, K. C. (2018). Internationalising smart specialisation. *Journal of the Knowledge Economy, 9*(1), 263–293.

Rodríguez-Pose, A. (2018). The revenge of the places that don't matter. *Cambridge Journal of Regions, Economy and Society, 11*(1), 189–209.

Rodríguez-Pose, A., Bartalucci, F., Lozano-Gracia, N., & Dávalos, M. (2024). Overcoming left-behindness. *Regional Science Policy & Practice, 16*(12), 100144.

Rodríguez-Pose, A., & Crescenzi, R. (2008). Research and development, spillovers, innovation systems, and the genesis of regional growth in Europe. *Regional Studies, 42*(1), 51–67.

Rodríguez-Pose, A., Dijkstra, L., & Poelman, H. (2024). The geography of EU discontent and the regional development trap. *Economic Geography, 100*(3), 213–245. doi:10.1080/00130095.2024.2337657.

Rosenberg, N. (1994). *Exploring the black box: Technology, economics, and history*. Cambridge University Press.

Scherngell, T., & Barber, M. J. (2009). Spatial interaction modelling of cross-region R&D collaborations: Empirical evidence from the 5th EU Framework Programme. *Papers in Regional Science, 88*(3), 531–546.

Sotarauta, M., & Pulkkinen, R. (2011). Institutional entrepreneurship for knowledge regions: In search of a fresh set of questions for regional innovation studies. *Environment and Planning C: Government and Policy, 29*(1), 96–112.

Szücs, F. (2018). Research subsidies, industry–university cooperation and innovation. *Research Policy, 47*(7), 1256–1266. doi:10.1016/j.respol.2018.04.009.

Tiffin, S., & Kunc, M. (2011). Measuring university roles in regional innovation. *Science and Public Policy, 38*(1), 55–66.

Uhlbach, W.-H., Balland, P.-A., & Scherngell, T. (2022). Public R&D funding and new regional specialisations: The contingent role of technological relatedness. *Industry and Innovation, 29*(4), 511–532. doi:10.1080/13662716.2022.2043147.

Uyarra, E., & Flanagan, K. (2010). From regional systems of innovation to regions as innovation policy spaces. *Environment and Planning C: Government and Policy, 28*(4), 681–695.

Uyarra, E., Marzocchi, C., & Sörvik, J. (2018). How outward looking is smart specialisation? *European Planning Studies*.

Varela-Vázquez, P., González-López, M., & Sánchez-Carreira, M. C. (2019). EU Framework Programmes distribution. *Journal of Entrepreneurship, Management and Innovation, 15*(3), 45–72.

Velthuis, S., Royer, J., Le Petit-Guerin, M., Cauchi-Duval, N., Franklin, R., Leibert, T., MacKinnon, D., & Pike, A. (2025). Regional varieties of left-behindness. *Regional Studies, 59*(1).

Wanzenböck, I., Neuländtner, M., & Scherngell, T. (2020). Impacts of EU funded R&D networks on the generation of key enabling technologies: Empirical evidence from a regional perspective. *Papers in Regional Science, 99*(1), 3–24. regionalscience.org

Wanzenböck, I., & Piribauer, P. (2018). R&D networks and regional knowledge production in Europe: Evidence from a space-time model. *Papers in Regional Science, 97*(S1), S1–S24. doi:10.1111/pirs.12236. Utrecht University

Wolfe, D. A. (2005). The role of universities in regional development. In *Creating knowledge, strengthening nations* (pp. 167–194).

Wooldridge, J. M. (2015). Control function methods in applied econometrics. *Journal of Human Resources, 50*(2), 420–445.

## Tables

*Table 1 - Descriptive statistics and frequency distribution*

### Descriptive statistics

| Variable | Source | Obs | Mean | SD | Min | Max |
|---|---|---|---|---|---|---|
| *Patents per capita* | OECD Regpat, ARDECO | 15158 | 0.67 | 0.85 | 0.00 | 4.25 |
| *Public R&D expenditures* | EUPRO, ARDECO | 15158 | 3.00 | 2.43 | 0.00 | 10.01 |
| *IV Public R&D expenditures* | EUPRO, ARDECO | 15158 | -3.80 | 8.98 | -15.58 | 14.54 |
| *GDP per capita* | ARDECO | 15158 | 3.79 | 0.63 | 1.76 | 5.91 |
| *Employment population ratio* | ARDECO | 15158 | 0.43 | 0.10 | 0.06 | 1.05 |
| *Investment labour ratio* | ARDECO | 15158 | 1.78 | 0.39 | 0.32 | 2.95 |
| *Patent stock per capita* | OECD Regpat, ARDECO | 15158 | 0.38 | 0.55 | 0.00 | 3.41 |
| *Share GVA Agriculture* | ARDECO | 15158 | 0.03 | 0.04 | 0.00 | 0.29 |
| *Share GVA Manufacturing* | ARDECO | 15158 | 0.23 | 0.10 | -0.02 | 0.75 |
| *Share GVA Construction* | ARDECO | 15158 | 0.06 | 0.02 | 0.00 | 0.21 |
| *Share GVA Market Services* | ARDECO | 15158 | 0.42 | 0.08 | 0.13 | 0.73 |
| *Share GVA Non Market Services* | ARDECO | 15158 | 0.23 | 0.07 | 0.05 | 0.59 |

### Frequency distributions

| Variable | Source | No | Yes |
|---|---|---|---|
| *Univ* | Authors' own computation | 8138 | 7020 |
| *DT risk* | ARDECO | 11891 | 3267 |
| *DT intensity* | ARDECO | 11193 | 3965 |
| *DT duration* | ARDECO | 12016 | 3142 |

*Table 2 - Correlation matrix*

| | *Patents per capita* | *Public R&D expenditures* | *University* | *DT risk* | *DT intensity* | *DT duration* | *IV Public R&D expenditures* | *GDP per capita* | *Employment population ratio* | *Investment labour ratio* | *Patent stock per capita* | *Share GVA Agriculture* | *Share GVA Manufacturing* | *Share GVA Construction* | *Share GVA Market Services* | *Share GVA Non Market Services* |
|---|---|---|---|---|---|---|---|---|---|---|---|---|---|---|---|---|
| *Patents per capita* | 1.00 | | | | | | | | | | | | | | | |
| *Public R&D expenditures* | 0.15 | 1.00 | | | | | | | | | | | | | | |
| *University* | 0.04 | 0.38 | 1.00 | | | | | | | | | | | | | |
| *DT risk* | 0.06 | 0.03 | 0.04 | 1.00 | | | | | | | | | | | | |
| *DT intensity* | -0.15 | -0.01 | 0.01 | -0.14 | 1.00 | | | | | | | | | | | |
| *DT duration* | 0.06 | 0.06 | 0.04 | 0.75 | -0.11 | 1.00 | | | | | | | | | | |
| *IV Public R&D expenditures* | -0.06 | -0.29 | -0.15 | -0.03 | -0.04 | -0.01 | 1.00 | | | | | | | | | |
| *GDP per capita* | 0.56 | 0.28 | 0.18 | -0.07 | -0.20 | -0.07 | -0.09 | 1.00 | | | | | | | | |
| *Employment population ratio* | 0.42 | 0.23 | 0.21 | 0.07 | -0.06 | 0.06 | -0.11 | 0.64 | 1.00 | | | | | | | |
| *Investment labour ratio* | 0.42 | 0.15 | 0.06 | -0.06 | -0.21 | -0.06 | -0.05 | 0.23 | 0.73 | 1.00 | | | | | | |
| *Patent stock per capita* | 0.97 | 0.22 | 0.03 | 0.05 | -0.13 | 0.05 | -0.08 | 0.41 | 0.54 | 0.39 | 1.00 | | | | | |
| *Share GVA Agriculture* | -0.49 | -0.27 | -0.19 | -0.04 | 0.15 | -0.02 | 0.09 | -0.50 | -0.66 | -0.42 | -0.47 | 1.00 | | | | |
| *Share GVA* | 0.22 | -0.09 | -0.14 | 0.11 | 0.03 | 0.11 | 0.00 | 0.16 | 0.00 | -0.05 | 0.22 | 0.01 | 1.00 | | | |
| *Manufacturing* | | | | | | | | | | | | | | | | |
| *Share GVA Construction* | -0.16 | -0.27 | -0.13 | 0.16 | -0.08 | 0.13 | 0.07 | -0.29 | -0.26 | -0.05 | -0.16 | 0.26 | -0.02 | 1.00 | | |
| *Share GVA Market Services* | 0.06 | 0.28 | 0.19 | 0.00 | -0.08 | 0.01 | -0.05 | 0.08 | 0.25 | 0.23 | 0.05 | -0.36 | -0.66 | -0.12 | 1.00 | |
| *Share GVA Non Market Services* | -0.13 | 0.00 | 0.11 | -0.23 | -0.09 | -0.23 | -0.02 | -0.18 | 0.02 | 0.02 | -0.13 | 0.00 | -0.57 | -0.02 | 0.04 | 1.00 |

*Table 3- Linear OLS Fixed effects model estimation results - University*

| VARIABLES | (1)<br>Patents | (2)<br>Patents | (3)<br>Patents | (5)<br>Patents |
|---|---|---|---|---|
| Public R&D expenditures | 0.0856*** | 0.0744*** | | |
| | (0.0258) | (0.0238) | | |
| Public R&D expenditures# No University | | | 0.0310 | 0.3249** |
| | | | (0.0354) | (0.1473) |
| Public R&D expenditures # University | | | 0.0765*** | 0.3580** |
| | | | (0.0246) | (0.1398) |
| GDP per capita | | 0.2069*** | 0.2068*** | 0.2176*** |
| | | (0.0614) | (0.0614) | (0.0617) |
| Employment population ratio | | -0.1010** | -0.1012** | -0.0774* |
| | | (0.0404) | (0.0404) | (0.0411) |
| Investment labour ratio | | -0.1146*** | -0.1147*** | -0.1288*** |
| | | (0.0406) | (0.0406) | (0.0415) |
| Patent stock per capita | | 0.0120 | 0.0119 | 0.0074 |
| | | (0.0075) | (0.0075) | (0.0073) |
| Share GVA Agriculture | | 0.0804*** | 0.0803*** | 0.0866*** |
| | | (0.0253) | (0.0253) | (0.0255) |
| Share GVA Manufacturing | | 0.0789*** | 0.0786*** | 0.1194*** |
| | | (0.0188) | (0.0188) | (0.0283) |
| Share GVA Construction | | -0.2401*** | -0.2412*** | -0.2157*** |
| | | (0.0450) | (0.0450) | (0.0450) |
| Share GVA Market Services | | -0.0022 | -0.0024 | -0.0011 |
| | | (0.0175) | (0.0175) | (0.0176) |
| Share GVA Non Market Services | | 0.1191*** | 0.1187*** | 0.1281*** |
| | | (0.0381) | (0.0381) | (0.0388) |
| Residuals | | | | -0.2858** |
| | | | | (0.1436) |
| Constant | 0.0001*** | 0.0001*** | 0.0001*** | 0.0001* |
| | (0.0000) | (0.0000) | (0.0000) | (0.0000) |
| Observations | 15,158 | 15,158 | 15,158 | 15,158 |
| F-test IV # No University | | | | 14.97*** |
| F-test IV # University | | | | 59.79*** |
| Linear combination test | | | | -0.0596 |
| | | | | (0.0433) |
| R-squared | 0.9364 | 0.9372 | 0.9372 | 0.9372 |

Robust standard errors in parentheses. *** p<0.01, ** p<0.05, * p<0.1

*Table 4 - Linear OLS Fixed effects model estimation results - University and Development trap*

| | Dep. Var.: *Patents per capita* | | | | | |
|---|---|---|---|---|---|---|
| Development trap type: | *Risk* | | *Intensity* | | *Duration* | |
| | (1) | (2) | (3) | (4) | (5) | (6) |
| Public R&D expenditures # No University #No Dev. Trap | 0.0107 | 0.2723* | -0.0272 | 0.2651* | -0.0130 | 0.2427 |
| | (0.0456) | (0.1571) | (0.0358) | (0.1473) | (0.0516) | (0.1570) |
| Public R&D expenditures # No University # Dev. trap | 0.0793 | 0.3366** | 0.1546* | 0.4479*** | 0.1007** | 0.3548** |
| | (0.0485) | (0.1582) | (0.0824) | (0.1645) | (0.0420) | (0.1592) |
| Public R&D expenditures # University #No Dev. Trap | 0.0401* | 0.2920* | 0.0913*** | 0.3720*** | 0.0354 | 0.2808* |
| | (0.0228) | (0.1497) | (0.0285) | (0.1396) | (0.0226) | (0.1487) |
| Public R&D expenditures # University # Dev. Trap | 0.1637** | 0.4064*** | -0.0124 | 0.2665* | 0.1715** | 0.4131*** |
| | (0.0689) | (0.1474) | (0.0304) | (0.1439) | (0.0667) | (0.1503) |
| GDP per capita | 0.1972*** | 0.2074*** | 0.2066*** | 0.2174*** | 0.1994*** | 0.2090*** |
| | (0.0620) | (0.0625) | (0.0616) | (0.0619) | (0.0616) | (0.0622) |
| Employment population ratio | -0.1043** | -0.0831** | -0.1021** | -0.0785* | -0.1020** | -0.0813** |
| | (0.0408) | (0.0415) | (0.0404) | (0.0412) | (0.0403) | (0.0411) |
| Investment labour ratio | -0.1160*** | -0.1284*** | -0.1168*** | -0.1310*** | -0.1173*** | -0.1295*** |
| | (0.0409) | (0.0419) | (0.0407) | (0.0415) | (0.0408) | (0.0419) |
| Patent stock per capita | 0.0114 | 0.0075 | 0.0116 | 0.0071 | 0.0113 | 0.0074 |
| | (0.0073) | (0.0071) | (0.0075) | (0.0073) | (0.0073) | (0.0072) |
| Share GVA Agriculture | 0.0814*** | 0.0869*** | 0.0813*** | 0.0876*** | 0.0823*** | 0.0877*** |
| | (0.0255) | (0.0257) | (0.0254) | (0.0256) | (0.0255) | (0.0258) |
| Share GVA Manufacturing | 0.0790*** | 0.1151*** | 0.0774*** | 0.1180*** | 0.0789*** | 0.1142*** |
| | (0.0188) | (0.0284) | (0.0188) | (0.0282) | (0.0188) | (0.0290) |
| Share GVA Construction | -0.2395*** | -0.2171*** | -0.2406*** | -0.2152*** | -0.2393*** | -0.2172*** |
| | (0.0448) | (0.0443) | (0.0450) | (0.0451) | (0.0448) | (0.0451) |
| Share GVA Market Services | -0.0055 | -0.0042 | -0.0039 | -0.0026 | -0.0037 | -0.0025 |
| | (0.0178) | (0.0179) | (0.0177) | (0.0177) | (0.0176) | (0.0177) |
| Share GVA Non Market Services | 0.1204*** | 0.1285*** | 0.1203*** | 0.1297*** | 0.1213*** | 0.1294*** |
| | (0.0382) | (0.0389) | (0.0381) | (0.0387) | (0.0382) | (0.0389) |
| Residuals | | -0.2540* | | -0.2847** | | -0.2483* |
| | | (0.1519) | | (0.1429) | | (0.1511) |
| Observations | 15,158 | 15,158 | 15,158 | 15,158 | 15,158 | 15,158 |
| F-test IV # No University # No Dev. trap | | 24.99*** | | 13.38*** | | 17.46*** |
| F-test IV # No University # Dev. trap | | 11.01*** | | 13.00** | | 16.38*** |
| F-test IV # University # No Dev. trap | | 27.66*** | | 42.27*** | | 41.27*** |
| F-test IV # University # Dev. trap | | 54.83*** | | 22.48*** | | 58.86*** |
| Linear combination test | | -0.7576*** | | -0.8209*** | | -0.7871*** |
| | | (0.2886) | | (0.2939) | | (0.3017) |
| R-squared | 0.9372 | 0.9372 | 0.9372 | 0.9372 | 0.9372 | 0.9372 |

Robust standard errors in parentheses. *** p<0.01, ** p<0.05, * p<0.1

## Appendix

To address the potential endogeneity of the funding, Table A1 reports the first stage of the control function approach of table 3 and 4. The first-stage results reveal a heterogeneity by university presence: the instrument is positively associated with the funding stock in regions without a university, but negatively associated with it in university-hosting. We interpret this as a substitution pattern, whereby regions with an already-strong research base and direct EU funding access experience a partial crowding-out of their own funding growth as their FP6-era network partners themselves attract more H2020 resources, while non-university regions instead see funding move in tandem with partner exposure.

Table A1: CF first stages

| | Dep. Var.: *Public R&D expenditures* | | | |
|---|---|---|---|---|
| Development trap type: | | *Risk* | *Intensity* | *Duration* |
| | (1) | (2) | (3) | (4) |
| IV Public R&D expenditures # No University | 0.0004*** | | | |
| | (0.0001) | | | |
| IV Public R&D expenditures # University | -0.0024*** | | | |
| | (0.0003) | | | |
| IV Public R&D expenditures # No University # No Development trap | | 0.0006*** | 0.0005*** | 0.0005*** |
| | | (0.0001) | (0.0001) | (0.0001) |
| IV Public R&D expenditures # No University # Development trap | | -0.0002 | 0.0004* | 0.0003 |
| | | (0.0003) | (0.0002) | (0.0002) |
| IV Public R&D expenditures # University # No Development trap | | -0.0018*** | -0.0025*** | -0.0020*** |
| | | (0.0003) | (0.0004) | (0.0003) |
| IV Public R&D expenditures # University # Development trap | | -0.0046*** | -0.0022*** | -0.0031*** |
| | | (0.0007) | (0.0005) | (0.0004) |
| GDP per capita | -0.0482** | -0.0584*** | -0.0482** | -0.0527*** |
| | (0.0192) | (0.0196) | (0.0192) | (0.0196) |
| Employment population ratio | -0.0699*** | -0.0736*** | -0.0699*** | -0.0707*** |
| | (0.0191) | (0.0190) | (0.0191) | (0.0190) |
| Investment labour ratio | 0.0458 | 0.0447 | 0.0453 | 0.0460 |
| | (0.0328) | (0.0326) | (0.0326) | (0.0326) |
| Patent stock per capita | 0.0156*** | 0.0155*** | 0.0156*** | 0.0155*** |
| | (0.0018) | (0.0018) | (0.0018) | (0.0018) |
| Share GVA Agriculture | -0.0224 | -0.0204 | -0.0214 | -0.0214 |
| | (0.0233) | (0.0232) | (0.0234) | (0.0233) |
| Share GVA Manufacturing | -0.1405*** | -0.1373*** | -0.1400*** | -0.1393*** |
| | (0.0117) | (0.0117) | (0.0117) | (0.0117) |
| Share GVA Construction | -0.0838*** | -0.0833*** | -0.0834*** | -0.0841*** |
| | (0.0249) | (0.0249) | (0.0250) | (0.0249) |
| Share GVA Market Services | -0.0109 | -0.0094 | -0.0108 | -0.0094 |
| | (0.0108) | (0.0108) | (0.0109) | (0.0109) |
| Share GVA Non Market Services | -0.0375 | -0.0309 | -0.0373 | -0.0359 |
| | (0.0244) | (0.0244) | (0.0247) | (0.0244) |
| Observations | 15,158 | 15,158 | 15,158 | 15,158 |
| R-squared | 0.6617 | 0.6630 | 0.6617 | 0.6622 |

Robust standard errors in parentheses. *** p<0.01, ** p<0.05, * p<0.1